\input{aipcheck}

\documentclass[
    ,final            
  ]
  {aipproc}

\layoutstyle{6x9}

\begin{document}

\vspace*{-2.2cm}
\begin{flushright}
BNL-NT-03/18 \\
DO-TH 03/14 \\
RBRC-330 \\
\end{flushright}

\vspace*{-0.4cm}

\title{Double-transverse spin asymmetries at 
NLO\footnote{Invited talk
presented by W. Vogelsang at the ``Conference on the Intersections
of Particle and Nuclear Physics (CIPANP 2003)'', 
New York City, May 19-24, 2003.}}

\vspace*{-0.6cm}

\author{A.\ Mukherjee}{
  address={Institut f{\"u}r Physik, Universit{\"a}t Dortmund,
D-44221 Dortmund, Germany}
}

\author{M.\ Stratmann}{
  address={Institut f{\"u}r Theoretische Physik, Universit{\"a}t Regensburg,
D-93040 Regensburg, Germany}
}

\author{W.\ Vogelsang}{
  address={RIKEN-BNL Research Center and Physics Department, 
Brookhaven National Laboratory,\\
Upton, New York 11973, U.S.A.}
}

\vspace*{-1mm}
\begin{abstract}
We report on a next-to-leading order QCD calculation of the cross section 
and the spin asymmetry for isolated large-$p_T$ prompt photon production 
in collisions of transversely polarized protons. Corresponding measurements
may be used at RHIC to determine the transversity parton distributions of 
the proton.
\end{abstract}

\maketitle

\vspace*{-0.4cm}
The partonic structure of spin-1/2 targets at the leading-twist level
is characterized by the unpolarized, longitudinally polarized, 
and transversely polarized distribution functions $f$, $\Delta f$, 
and $\delta f$, respectively \cite{ref:jaffeji,ref:ralston}. 
These non-perturbative parton densities can be probed universally in 
a multitude of inelastic scattering processes, for which it is possible
to separate (``factorize'') the long-distance physics relating
to nucleon structure from a partonic short-distance scattering
that is amenable to QCD perturbation theory. 

In contrast to the $f$ and $\Delta f$, the ``transversity'' distributions
$\delta f$ are unmeasured thus far. They are presently the focus of much
experimental activity. For example, information should soon be gathered 
from transversely polarized proton-proton collisions at the BNL 
Relativistic Heavy Ion Collider (RHIC) \cite{ref:rhic}. 
The potential of RHIC in accessing transversity in measurements of 
transverse double-spin asymmetries $A_{\mathrm{TT}}$ 
was examined in \cite{ref:attlo} 
for high transverse momentum $p_T$ prompt photon and jet production
(for earlier studies, see \cite{ref:attold,ref:artru,ref:jaffesaito}). 
All of these calculations were performed only at the lowest order (LO) 
approximation for the underlying partonic hard-scattering.
As is well known, next-to-leading order (NLO) QCD 
corrections are generally indispensable in order to arrive at a firmer 
theoretical prediction for hadronic cross sections and spin asymmetries.
The NLO calculation for $A_{\mathrm{TT}}^{\gamma}$ 
for isolated high-$p_T$ prompt 
photon production, $pp\to\gamma X$, was recently completed~\cite{ref:msv};
here we give a brief report on those results. 

Interesting new technical questions arise beyond the LO in case of transverse
polarization. Unlike for longitudinally polarized cross sections where
the spin vectors are aligned with momentum, transverse spin vectors
specify extra spatial directions, giving rise to non-trivial
dependence of the cross section on the azimuthal angle of
the observed photon. As is well-known \cite{ref:ralston}, 
for $A_{\mathrm{TT}}$ this 
dependence is always of the form $\cos(2\Phi)$, if the $z$ axis is defined
by the direction of the initial protons in their center-of-mass
system (c.m.s.), and the spin vectors are 
taken to point in the $\pm x$ direction. Integration over 
the photon's azimuthal angle is therefore not appropriate. 
On the other hand, standard techniques developed in the literature 
for performing NLO phase-space integrations usually rely on 
integration over the full azimuthal phase space, and also on the 
choice of particular reference frames that are related in complicated 
ways to the one just specified. In~\cite{ref:msv},
a new general technique was introduced
which facilitates NLO calculations with transverse
polarization by conveniently projecting on the azimuthal dependence 
of the matrix elements in a covariant way.  The key point here
is to recognize that the the factor $\cos (2\Phi)$ in the cross 
section actually results from the covariant expression
\begin{equation} 
\label{eq5}
{\cal F}(p_{\gamma},s_a,s_b)\;=\;
\frac{s}{t u} 
\,\left[ 2 \,(p_{\gamma}\cdot s_a)\, (p_{\gamma}\cdot s_b)\; +\; 
\frac{t u}{s} \,(s_a \cdot s_b) \right] \;,
\end{equation} 
with $s_a, s_b$ the initial spin vectors and $p_{\gamma}$ the
photon momentum.  ${\cal F}$ 
reduces to $\cos (2\Phi)$ in the hadronic c.m.s.\ frame.
One may thus integrate over all phase space without obtaining 
a vanishing result if one simply multiplies the squared matrix element
by the factor ${\cal F}(p_{\gamma},s_a,s_b)$. Integration over
terms involving the $s_a,s_b$ can be carried out in a covariant
way by using standard tensor decompositions. 
After this step, there are no scalar products involving the $s_i$ left
in the squared matrix element. For the ensuing integration 
over all azimuthal phase space we can now employ techniques
familiar from the corresponding calculations in the unpolarized
and longitudinally polarized cases. 

At NLO, there are two subprocesses that contribute for transverse 
polarization, $q\bar{q}\to \gamma X$ and $qq \to \gamma X$.
The first one is already present at LO, where 
$X=g$. At NLO, one has virtual corrections to the Born
cross section ($X=g$), but also $2\to 3$ real emission
diagrams, with $X=gg+q\bar{q}+q'\bar{q}'$. For the second 
subprocess, $X=qq$. 

Owing to the presence of ultraviolet, infrared, and
collinear singularities at intermediate stages of the
calculation, it is necessary to introduce a regularization.
Our choice is dimensional regularization, that is, the 
calculation is performed in $d=4-2\varepsilon$ space-time 
dimensions. 
Ultraviolet poles in the virtual diagrams are removed by the 
renormalization of the strong coupling constant.
Infrared singularities cancel in the sum between virtual
and real-emission diagrams. After this cancelation,
only collinear poles are left. These result for example from a 
parton in the initial state splitting collinearly into 
a pair of partons, corresponding to a long-distance contribution in the 
partonic cross section. From the factorization theorem it follows that 
such contributions need to be factored into the parton distribution 
functions. 
In our calculations \cite{ref:msv}, we have imposed on the photon 
the isolation cut proposed in \cite{ref:frixione}. All 
{\it final-state} collinear singularities then cancel. The isolation 
constraint was implemented analytically by assuming a narrow isolation cone.

\begin{figure}
  \includegraphics[height=.45\textheight]{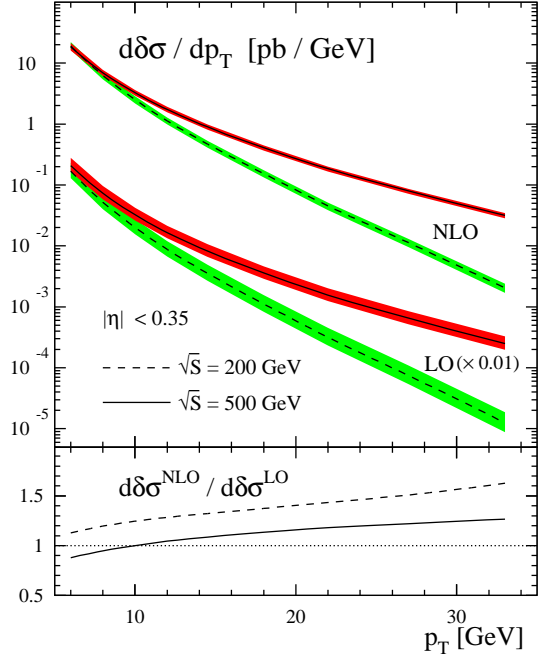}
  \caption{Predictions for the transversely polarized prompt 
photon production cross sections at LO and NLO, for $\sqrt{S}=200$ 
and 500 GeV. The LO results have been scaled by a factor of 0.01.
The shaded bands represent the theoretical uncertainty if
$\mu_F$ $(=\mu_R)$ is varied in the range $p_T/2\le \mu_F \le 2p_T$.
The lower panel shows the ratios of the NLO and LO results for both
c.m.s.\ energies. \label{fig:sigma}}
\end{figure}

For our numerical predictions we model the $\delta f$ by assuming
that the Soffer inequality \cite{ref:soffer} is saturated 
at some low input scale $\mu_0\simeq 0.6\,\mathrm{GeV}$.
For $\mu>\mu_0$ the transversity densities $\delta f(x,\mu)$ 
are then obtained by solving the appropriate QCD evolution equations. 
Our numerical predictions apply for prompt photon measurements 
with the {\sc Phenix} detector at RHIC. 
Figure~\ref{fig:sigma} shows our results for the transversely
polarized prompt photon production cross sections at NLO and LO
for two different c.m.s.\ energies.
The lower part of the figure displays the so called ``$K$-factor'',
$K=d\delta\sigma^{\rm NLO}/d\delta\sigma^{\rm LO}$.
One can see that the NLO corrections are somewhat smaller for
$\sqrt{S}=500\,\mathrm{GeV}$ and increase with $p_T$.
The shaded bands in the upper panel of Fig.~\ref{fig:sigma}
indicate the uncertainties from varying the factorization and 
renormalization scales in the range $p_T/2 \leq \mu_R=\mu_F \leq 2 p_T$. 
The solid and dashed lines are always for the choice where all scales are 
set to $p_T$, and so is the $K$ factor underneath. One can see that the scale 
dependence becomes much weaker at NLO, as expected. The corresponding
spin asymmetries $A_{\mathrm{TT}}^{\gamma}=d\delta\sigma/d\sigma$
may be found in Fig.~2 of Ref. \cite{ref:msv}; they are generally
smaller at NLO than at LO.

\begin{theacknowledgments}
W.V.\ is grateful to RIKEN, Brookhaven National Laboratory and the U.S.\
Department of Energy (contract number DE-AC02-98CH10886) for
providing the facilities essential for the completion of this work.
This work is supported in part by the ``Bundesministerium f\"{u}r Bildung und
Forschung (BMBF)'' and the ``Deutsche Forschungsgemeinschaft (DFG)''.
\end{theacknowledgments}


\bibliographystyle{aipproc}   

%

\end{document}